\documentclass[aip,jcp,reprint]{revtex4-1}
\usepackage{graphicx}

\begin{document}

\title{Methane and carbon dioxide adsorption on edge-functionalized graphene: A comparative DFT study} 

\author{Brandon C. Wood}
\affiliation{Quantum Simulations Group, Lawrence Livermore National Laboratory, Livermore, CA 94550 USA}
\affiliation{Theoretical Sciences Unit, Jawaharlal Nehru Centre for Advanced Scientific Research, Jakkur, Bangalore 560064 India}
\author{Shreyas Y. Bhide}
\affiliation{Department of Chemical Engineering, Indian Institute of Science, Bangalore 560012 India}
\author{Debosruti Dutta}
\affiliation{Department of Chemical Engineering, Indian Institute of Science, Bangalore 560012 India}
\author{Vinay S. Kandagal}
\affiliation{Department of Chemical Engineering, Indian Institute of Science, Bangalore 560012 India}
\author{Amar Deep Pathak}
\affiliation{Department of Chemical Engineering, Indian Institute of Science, Bangalore 560012 India}
\author{Sudeep N. Punnathanam}
\affiliation{Department of Chemical Engineering, Indian Institute of Science, Bangalore 560012 India}
\author{K. G. Ayappa}
\affiliation{Department of Chemical Engineering, Indian Institute of Science, Bangalore 560012 India}
\author{Shobhana Narasimhan}
\affiliation{Theoretical Sciences Unit, Jawaharlal Nehru Centre for Advanced Scientific Research, Jakkur, Bangalore 560064 India}

\begin{abstract}
With a view towards optimizing gas storage and separation in crystalline and disordered nanoporous carbon-based materials, we use {\it ab initio} density functional theory calculations to explore the effect of chemical functionalization on gas binding to exposed edges within model carbon nanostructures. We test the geometry, energetics, and charge distribution of in-plane and out-of-plane binding of CO$_2$ and CH$_4$ to model zigzag graphene nanoribbons edge-functionalized with COOH, OH, NH$_2$, H$_2$PO$_3$, NO$_2$, and CH$_3$. Although different choices for the exchange-correlation functional lead to a spread of values for the binding energy, trends across the functional groups are largely preserved for each choice, as are the final orientations of the adsorbed gas molecules. We find binding of CO$_2$ to exceed that of CH$_4$ by roughly a factor of two. However, the two gases follow very similar trends with changes in the attached functional group, despite different molecular symmetries. Our results indicate that the presence of NH$_2$, H$_2$PO$_3$, NO$_2$, and COOH functional groups can significantly enhance gas binding with respect to a hydrogen-passivated edge, making the edges potentially viable binding sites in materials with high concentrations of edge carbons. To first order, in-plane binding strength correlates with the larger permanent and induced dipole moments on these groups. Implications for tailoring carbon structures for increased gas uptake and improved CO$_2$/CH$_4$ selectivity are discussed.

\end{abstract}

\maketitle

\section{Background}

Nanoporous carbon-based materials are particularly attractive as gas storage media, due to their high intrinsic adsorptive capacity, low cost and weight, and relative ease of processing. Accordingly, both crystalline and disordered nanoporous carbons have been the subject of active investigation for vehicular storage of natural gas and hydrogen, as well as framework materials for CO$_2$ capture. \cite{celzard05, menon98, kabbour06, castello02a, sircar96, pevida08, casa-lillo02, strobel06, olajire10, wang11, furukawa09, biener11, james03, millward05} Among the adjustable processing parameters is the ability to tune the internal chemistry via insertion of desired functional groups to enhance the interaction between the substrate and gas adsorbate or increase the active surface area of the material, thereby increasing uptake. If the substrate-adsorbate interaction can be enhanced selectively, the host material may also be used for industrial processes such as natural gas purification, where efficient separation of CO$_2$ contaminants from CH$_4$ is necessary. \cite{dalessandro10, li09, bae08} In disordered porous materials, such as activated carbons and carbon aerogels, \cite{kabbour06, biener11, biener12} this is often achieved by thermal activation, pyrolysis, or chemical/electrochemical treatment. Depending on the specific nature of the chemical process employed, this can result in a predominance of certain classes of functional groups. \cite{chen04, shafeeyan10} The intrinsic disorder gives rise to significant local deviations from the dominant planar $sp^2$ geometry, including reasonable concentrations of reactive edges that act as ready binding sites for additional functional groups. \cite{brennan01, marsh06, boehm94, asai11, biener12} Efforts towards increasing gas uptake and selective separation have often focused on understanding the influence of activation on the resulting pore size distribution and available surface area. Generally speaking, less attention has been paid to the specific interactions of the functional groups with the adsorbate. This is in part due to complexities in characterizing which groups are present, as well as in separating the effects of increased surface area and topological defects from those of chemical functionalization. \cite{feng06} 

Crystalline porous framework materials, such as metal-organic frameworks (MOFs) and covalent-organic frameworks (COFs), offer additional tight control over the local chemistry within the framework itself. This is achieved by direct substitution or functionalization of molecular building blocks, which commonly feature segments of interconnected aromatic rings. \cite{xiang10, furukawa09, james03, cote05, wilmer12, deng10}

For both crystalline and disordered nanoporous carbon substrates, it is highly desirable to devise specific, well-founded strategies for modifying gas uptake via chemical functionalization or substitution. However, doing so reliably requires a robust understanding of how changes in the identity and local geometry of functional groups attached to edges of $sp^2$ carbon can affect the interaction with adsorbed gas molecules. Accordingly, we have used {\it ab initio} density functional theory to systematically study the influence of certain commonly occurring edge functional groups~\cite{boehm94, brennan01, figueiredo99} on methane and carbon dioxide adsorption within model carbon structures. The substrates are modeled as edge-functionalized graphene nanoribbons, which represent the simplest extended $sp^2$ carbon system with exposed edges, and as such are suitable for exploring edge functionalization in locally planar carbon structures.\cite{geim09, dutta10} Our results indicate that it is possible to activate exposed edges to act as additional gas binding sites, thereby increasing the active surface area and total gas uptake of the material. Given current interest towards more accurate theoretical treatments of the weak dispersion forces that dominate gas binding on these substrates, we also provide a careful comparison of values obtained using recently developed exchange-correlation functionals. We conclude that in most cases, trends are well preserved across competing methods.

\section{Computational Details}

We use density functional theory (DFT) to determine the geometry and energetics of CH$_4$ and CO$_2$ adsorbed on zigzag graphene nanoribbons (ZGNRs) with attached chemical functional groups. We test three electron-donating groups (OH, NH$_2$, CH$_3$) and three electron-withdrawing groups (NO$_2$, COOH, H$_2$PO$_3$). L\"owdin population (LP) analysis is employed to analyze the charge redistribution resulting from adsorption. We have restricted ourselves to considering chemical modification of only the zigzag edges of graphene nanoribbons; this choice is motivated by the results of earlier studies showing that zigzag edges are more reactive than armchair edges, and are therefore more likely binding sides for edge functional groups.\cite{kobayashi05, koskinen08} The reactivity of the zigzag edges was verified in our own test calculations, which indicated that unpassivated ZGNR edges can react spontaneously with nearby gas adsorbates. Similar behavior was reported in Ref.~\onlinecite{huang08}.
 
As physisorption dominates the interactions between the adsorbate and substrate, it is important to select a method of treating van der Waals (vdW) dispersion forces within a DFT framework that will qualitatively preserve the trends we are attempting to describe. This remains an area of active research.\cite{langreth09} For methane adsorption, we compare three choices for the exchange-correlation functional: 1) the uncorrected Perdew-Zunger local density approximation (DFT-LDA);\cite{perdew81} 2) the local atomic potential formulation (DFT+LAP); \cite{dftlap} and 3) the van der Waals density functional (vdW-DF). \cite{langreth09, dion04, thonhauser07} The DFT+LAP and vdW-DF functionals are specifically designed to reproduce accurate binding energies for physisorbed systems by taking into account van der Waals dispersion forces. 

The DFT+LAP functional is based on the augmentation of an existent pseudopotential by an additional channel, expressed in an atom-centered local form.\cite{dftlap} The correction is constant for short-range interactions but falls off exponentially for long-range interactions. The three independent fitting parameters in DFT+LAP are derived from binding curves of small dimer systems, as obtained from coupled-cluster calculations including single, double, and perturbative triple excitations (CCSD(T)). Here, we apply the correction to pseudopotentials generated using the revPBE functional.\cite{revpbe} It should be noted that DFT+LAP values are not given for the H$_2$PO$_3$-functionalized ZGNR due to the unavailability of a parameterized local correction for phosphorus. 

The vdW-DF functional mixes short-range LDA correlation with a fully nonlocal, \emph{ab initio} approximation of long-range correlation, which contains no empirical parameters.\cite{langreth09} In principle, the correction may be implemented self-consistently. However, numerous calculations of weak-binding molecular systems have shown that non-self consistent perturbation of a properly chosen exchange functional is usually sufficient, resulting in substantial computational savings.\cite{langreth09,li08}. Accordingly, we adopt the non-self consistent formulation of vdW-DF here. As with DFT+LAP, we have used the revPBE functional~\cite{revpbe} to compute the reference exchange term. 

All DFT calculations were performed using the Quantum-ESPRESSO code. \cite{espresso} For the DFT-LDA and DFT+LAP calculations, we used ultrasoft pseudopotentials \cite{vanderbilt90, rappe90} with plane-wave cutoffs of 30~Ry and 360~Ry for the wavefunctions and charge density, respectively. For the vdW-DF calculations, norm-conserving pseudopotentials with a plane-wave cutoff of 80 Ry were used. For the periodic ribbon calculations, a vacuum spacing of 20 \AA\ was introduced along non-repeating directions, and a $6 \times 1 \times 1$ Monkhorst-Pack $\mathbf{k}$-point mesh was used. Marzari-Vanderbilt cold smearing \cite{marzari99} with a width of 0.007~Ry was introduced for improved convergence and for savings in $\mathbf{k}$-point density. In most cases, calculations were performed with unrestricted spin; however, for the vdW-DF calculations, spin-restricted DFT was used, since the methodology is not rigorously defined for the spin-unrestricted case. In the spin-restricted calculations, all functionalized ZGNRs demonstrate semiconducting or half-semiconducting behavior, in agreement with previous theoretical work on narrow ribbons.\cite{li08,hod07,son06} In addition, we observe that the ground-state ZGNR spin configuration consists of two sets of ferromagnetically aligned edge states which are oppositely oriented at opposite ribbon edges, as has been reported elsewhere. \cite{son06,fujita96}

For comparison with the DFT adsorption numbers on the functionalized ZGNRs, second-order M{\o}ller-Plesset perturbation theory (MP2) adsorption calculations were performed for methane on similarly functionalized benzene rings. MP2 calculations are known to capture correlation effects, including van der Waals dispersion. We used the Dunning aug-cc-pVDZ basis set \cite{dunning} to describe all atoms, with the counterpoise correction method \cite{bsse} employed to account for basis set superposition error. MP2 calculations were performed using the Gaussian09 software.\cite{gaussian} Isolated benzene rings were substituted for the ZGNRs in the MP2 calculations in order to make the calculations computationally accessible. Although the lower coordination of the benzene rings results in weaker overall adsorption and precludes direct quantitative comparison against our ZGNR values, the MP2 binding trends across the functional groups are useful as an additional benchmark. They also aid in forming a qualitative assessment of the relative contribution of the functional group versus the ribbon in determining adsorption behavior. 

\begin{figure}
\centering
\includegraphics[width=2.5in]{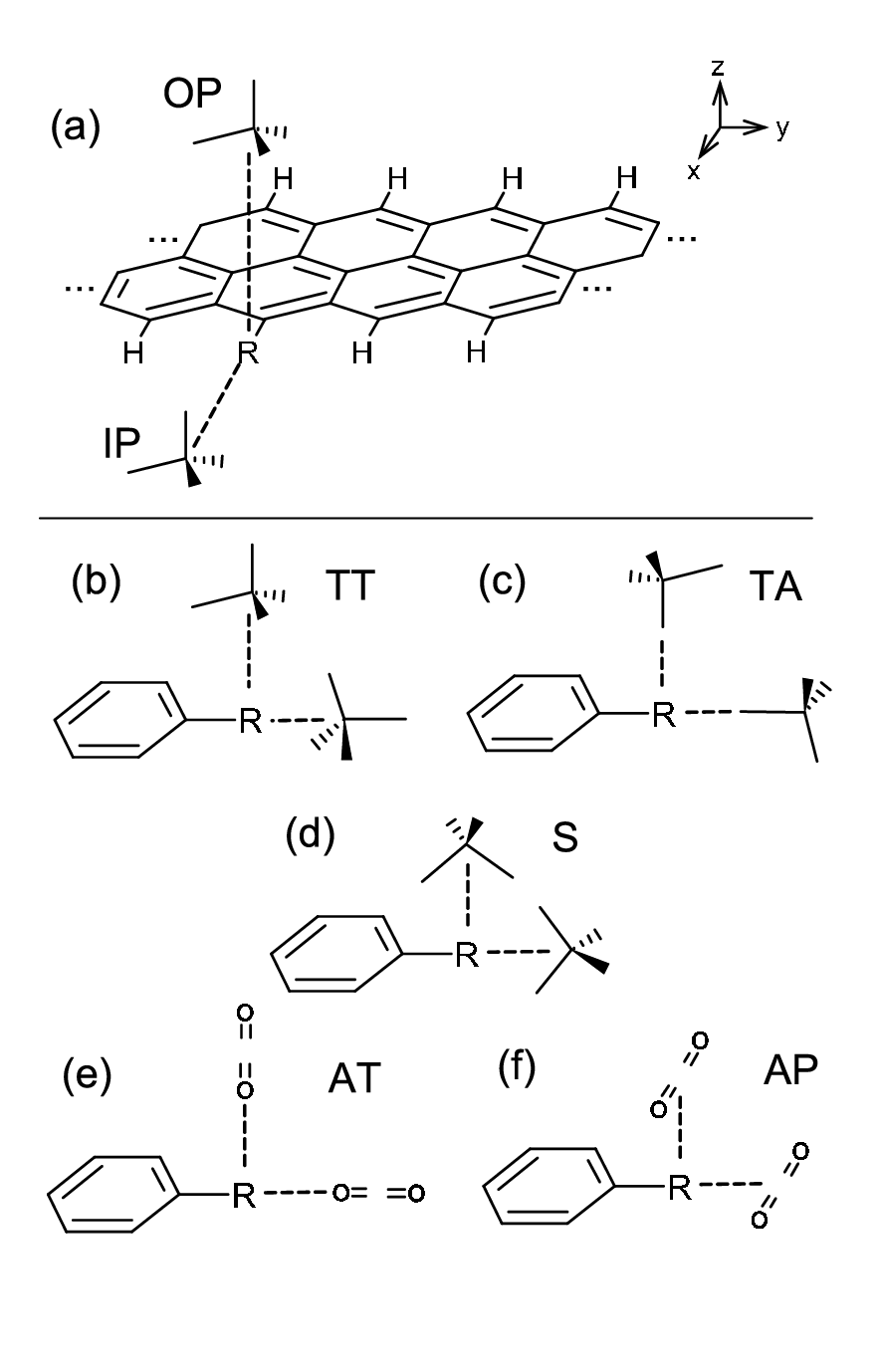}
\caption{Schematic representation of the tested adsorption geometries listed in Tables~\ref{Tab:eads_ch4} and \ref{Tab:eads_co2} and in the text. Here, $R$ represents the functional-group adsorption site. Panels illustrate (a) in-plane (IP) and out-of-plane (OP) adsorption; (b) CH$_4$ adsorption oriented with the tripod towards (TT) the functional group; (c) CH$_4$ adsorption oriented with the tripod away (TA) from the functional group; (d) CH$_4$ adsorption oriented such that two C--H bonds straddle (S) the functional group; (e) CO$_2$ adsorption oriented with the C=O axis toward (AT) the functional group; and (f) CO$_2$ adsorption oriented with the C=O axis perpendicular (AP) to the functional group.}
\label{bindgeom}
\end{figure}

The model ZGNRs used in our calculations are six carbon atoms wide with a single functional group attached to a carbon along one zigzag edge. Since we are interested in dilute functionalization, we include in the supercell three additional zigzag-edge carbons adjacent to the functional group, resulting in a total of 24 carbon atoms (see Fig.~\ref{bindgeom}).  These additional carbons are terminated with hydrogen, as are all dangling bonds on the ribbon edge opposite to the functional group. As a reference, gas adsorption is also studied on a fully hydrogen-passivated ZGNR. The benzene molecules studied using MP2 also have a single attached functional group, with hydrogens attached to the remaining carbons in the ring. The adsorption energy is defined as $ E_\mathrm{ads} = - E_\mathrm{g+s} + E_\mathrm{g} + E_\mathrm{s}$, where $E_\mathrm{g+s}$ is the energy of the adsorbed system, $E_\mathrm{g}$ is the energy of an isolated gas molecule (CH$_4$ or CO$_2$), and $E_\mathrm{s}$ is the energy of the substrate alone. 

In order to better span the configurational phase space of possible CH$_4$ and CO$_2$ geometries, we examine two different initial adsorbate binding sites for each complex, from which the system is allowed to relax without constraint. These are shown in Fig.~\ref{bindgeom}(a). In the first, which we refer to as ``in-plane'' (IP) adsorption, the CH$_4$ or CO$_2$ molecule is initially placed adjacent to the functional group, with the carbon atom of the gas molecule lying in the basal plane. In the second, referred to as ``out-of-plane'' (OP) adsorption, the gas molecule is initially placed directly above the functional group, perpendicular to the basal plane. These two configurations were selected based on symmetry considerations so as to maximize (OP) or minimize (IP) interaction with the ZGNR or benzene simultaneously with the functional group. Notably, these same sites were found as the two local extrema in a recent comprehensive study of the potential energy surface of water binding to benzene, further motivating their choice here. \cite{li08, langreth09}

Three possible high-symmetry orientations exist for bound methane, depicted schematically in Fig.~\ref{bindgeom}(b). In the first, which we label ``tripod toward'' (TT), one of the methane C--H bonds along the C$_{3}$ molecular axis points toward the functional group. In the second, ``tripod away'' (TA), one of the C--H bonds along the C$_{3}$ axis points away from the functional group. In the third, ``straddle'' (S), the C$_{2}$ axis aligns with the functional group. Similarly, CO$_2$ can be found in two high-symmetry orientations, shown in Fig.\ref{bindgeom}(c). The first, labeled ``axis toward'' (AT), has the $C_\infty$ axis of CO$_2$ aligned with the functional group. The second, labeled ``axis perpendicular'' (AP), instead has the $C_2$ axis of the CO$_2$ aligned with the functional group. Each of these was tried as the starting geometry, although we report only the lowest-energy final configurations here. When recording the final configurations for OP binding, we also indicate the approximate final location of the adsorbed gas molecule with respect to the basal plane, either above one of the functional group atoms (G) or else above the carbons on the ribbon edge (E).

Given the difficulty of controlled synthesis of edge-functionalized graphene, we cannot directly benchmark our nanoribbon results against experiments. Accordingly, we have also calculated OP binding of CH$_4$ and CO$_2$ to clean graphene sheets in a $4\times4$ unit cell, which allows for a more straightforward comparison. Our values of 12.1 kJ/mol for DFT-LDA and 11.1 kJ/mol for DFT+LAP compare favorably with the experimental value of 12.2 kJ/mol for methane on graphite, \cite{vidali91} and match well with previous DFT-LDA calculations.\cite{yang06} The effect on gas binding of adding layers to single-layer graphene was found to be negligibly small, both in our test calculations and in previous studies.\cite{yang06,thierfelder11} It is worth noting that DFT-LDA reproduces experimental methane binding on graphite quite well without correction, in agreement with previous analyses of similar systems in the literature.\cite{girifalco02, lu09, okamoto01, zhao02} Our OP vdW-DF value of 16.9 kJ/mol is slightly larger than the value of 15.4 kJ/mol in Ref.~\onlinecite{thierfelder11}, although both numbers are higher than the experimental value, in keeping with reports of overestimation of vdW-DF binding on $\pi$ systems.\cite{thonhauser07} We also benchmarked our MP2 calculations against available CCSD(T) calculations for methane binding above the ring centers of benzene and phenol;\cite{ringer06} our calculated binding energies of 6.4 kJ/mol and 6.5 kJ/mol for benzene and phenol, respectively, are similar to but slightly higher than the CCSD(T) values of 5.9 and 6.2 in the complete basis set limit. This is consistent with similar analyses in the literature. \cite{ringer06, uchimaru00}

\section{Methane adsorption}

\begin{figure*}
\centering
\includegraphics[width=5.0in]{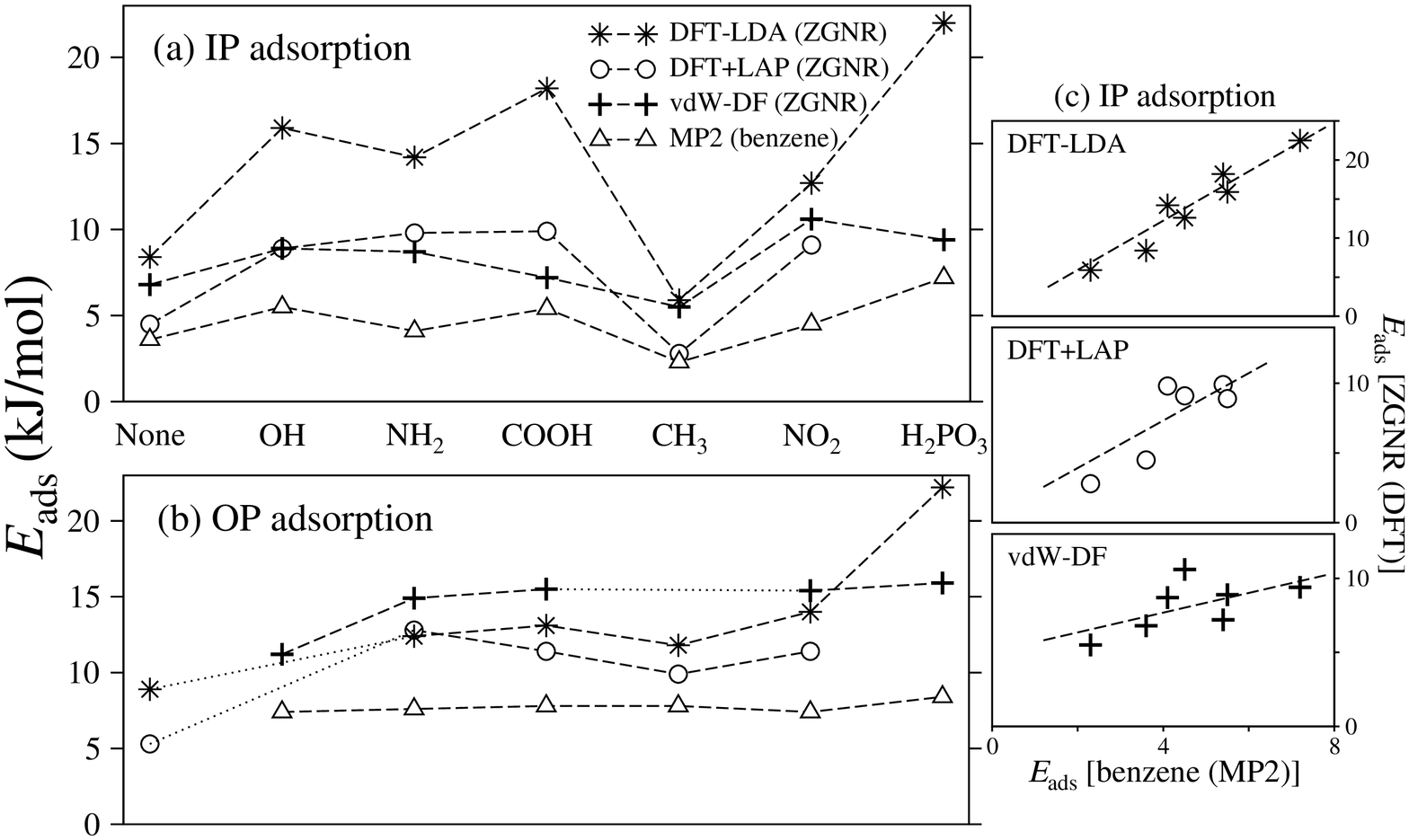}
\caption{Trend in calculated adsorption energies $E_\mathrm{ads}$ for (a) IP and (b) OP binding of methane on ZGNRs across various functional groups using DFT-LDA (stars), DFT+LAP (circles), and vdW-DF (crosses) exchange-correlation functionals. `None' refers to the fully hydrogen-passivated ZGNR. Adsorption energies are also shown for MP2 calculations (triangles) of methane binding to similarly functionalized benzene rings. Panel (c) shows the correlation of the DFT results on the ZGNRs with the MP2 results on benzene for IP binding.
}
\label{fig:eads_bymethod}
\end{figure*}

Adsorption energies for IP and OP methane binding to the functionalized ZGNRs are given in Fig.~\ref{fig:eads_bymethod} and Table~\ref{Tab:eads_ch4} for three different DFT exchange-correlation functionals: DFT-LDA, DFT+LAP, and vdW-DF, as well as MP2 results for methane on similarly functionalized benzene rings. It is immediately obvious from Fig.~\ref{fig:eads_bymethod} and Table~\ref{Tab:eads_ch4} that the different techniques can lead to a wide range of values for $E_\mathrm{ads}$. This is particularly true for IP adsorption, where DFT-LDA overbinds by about a factor of two with respect to the van der Waals-corrected DFT methods. For OP adsorption, the energies for the ZGNRs are much closer, usually falling within 30\% of one another. As expected, addition of van der Waals corrections to DFT decreases $E_\mathrm{ads}$ for IP binding, compensating the overbinding within DFT-LDA. However, in the case of vdW-DF, these same corrections increase $E_\mathrm{ads}$ relative to DFT-LDA for OP binding. Note that vdW-DF has been shown to overbind $\pi$ systems,\cite{thonhauser07} which may explain why it gives larger values for OP $E_\mathrm{ads}$ compared with the other methods. For both geometries, the MP2-derived $E_\mathrm{ads}$ is significantly weaker, reflecting the lower coordination of the benzene system. 

\begin{table*}
\centering
\caption{Final relaxed geometries and adsorption energies $E_{\mathrm{ads}}$ of CH$_4$ adsorbed on functionalized and unfunctionalized ZGNRs. The final orientation and position of the CH$_4$ molecule is given alongside the adsorption energy in parentheses (see text for description of acronyms). For IP adsorption, $d_\mathrm{min}$ represents the distance from the CH$_4$ carbon atom to the nearest functional-group atom, whose identity is given in parentheses. For OP adsorption, $d_z$ is the distance of the adsorbate from the ribbon basal plane. ``Unstable'' refers to configurations for which no stable minimum was found.}
\begin{tabular}
        {ccccccc}
        \hline\hline
        Group &  Site & Quantity & DFT-LDA & DFT+LAP & vdW-DF & MP2 (benzene) \\
        \hline
         None (H) &  IP & $E_{\mathrm{ads}}$ (kJ/mol) & 8.4 (TT) & 4.5 (TT) & 6.8 (TT) & 3.6 (TT) \\
                        &  & $d_{\mathrm{min}}$ (\AA)           & 2.59 (H) & 2.85 (H) & 3.03 (H) & 3.05 (H) \\
                        & OP & $E_{\mathrm{ads}}$ (kJ/mol) & 8.9 (TT,G) & 5.3 (TT,G) & Unstable & Unstable \\
                        &  & $d_z$ (\AA)                                 & 2.67 & 3.05 & Unstable & Unstable \\
        \hline
         OH          &  IP & $E_{\mathrm{ads}}$ (kJ/mol) & 15.9 (TT) & 8.9 (TT) & 8.9 (TT) & 5.5 (S) \\
                        &  & $d_{\mathrm{min}}$ (\AA)           & 2.23 (H) & 2.51 (H) & 2.89 (H) & 2.58 (H) \\
                        & OP & $E_{\mathrm{ads}}$ (kJ/mol) & Unstable & Unstable & 11.2 (TT,E) & 7.4 (TT,E) \\
                        &  & $d_z$ (\AA)                                 & Unstable & Unstable & 3.56 & 3.56 \\
        \hline
         NH$_2$   &  IP & $E_{\mathrm{ads}}$ (kJ/mol) & 14.2 (S) & 9.8 (TT) & 8.7 (TT) & 4.0 (TT) \\
                        &  & $d_{\mathrm{min}}$ (\AA)           & 2.47 (H) & 2.64 (H) & 3.11 (H) & 2.93 (H) \\
                        & OP & $E_{\mathrm{ads}}$ (kJ/mol) & 12.4 (TT,E) & 12.8 (TT,E) & 14.9 (TT,E) & 7.6 (S,E) \\
                        &  & $d_z$ (\AA)                                 & 3.18 & 3.51 & 3.72 & 3.53 \\
        \hline
       COOH       &  IP & $E_{\mathrm{ads}}$ (kJ/mol) & 18.2 (TT) & 9.9 (TT) & 7.2 (TT) & 5.4 (TT) \\
                        &  & $d_{\mathrm{min}}$ (\AA)           & 2.08 (H) & 2.36 (H) & 2.70 (H) & 2.58 (H) \\
                        & OP & $E_{\mathrm{ads}}$ (kJ/mol) & 13.1 (TT,E) & 11.4 (TT,E) & 15.5 (TT,E) & 7.8 (TT,E) \\
                        &  & $d_z$ (\AA)                                 & 3.20 & 3.43 & 3.72 & 3.46 \\
        \hline
       CH$_3$     &  IP & $E_{\mathrm{ads}}$ (kJ/mol) & 5.9 (TT) & 2.8 (TT) & 5.5 (TT) & 2.3 (TT) \\
                        &  & $d_{\mathrm{min}}$ (\AA)           & 2.95 (H) & 2.99 (H) & 3.47 (H) & 3.47 (H) \\
                        & OP & $E_{\mathrm{ads}}$ (kJ/mol) & 11.8 (TT,E) & 9.9 (TT,E) & Unstable & 7.8 (TT,E) \\
                        &  & $d_z$ (\AA)                                 & 3.36 & 3.56 & Unstable & 3.65 \\
        \hline
       NO$_2$    &  IP & $E_{\mathrm{ads}}$ (kJ/mol) & 12.7 (S) & 9.1 (S) & 10.6 (S) & 4.5 (S) \\
                        &  & $d_{\mathrm{min}}$ (\AA)           & 3.28 (O) & 3.57 (O) & 3.92 (O) & 3.97 (O) \\
                        & OP & $E_{\mathrm{ads}}$ (kJ/mol) & 14.0 (TT,E) & 11.4 (TT,E) & 15.4 (TT,E) & 7.4 (S,E) \\
                        &  & $d_z$ (\AA)                                 & 3.22 & 3.45 & 3.72 & 3.55 \\
        \hline
H$_2$PO$_3$ &  IP & $E_{\mathrm{ads}}$ (kJ/mol) & 22.0 (TT) & N/A & 9.4 (TT) & 7.2 (TT) \\
                        &  & $d_{\mathrm{min}}$ (\AA)           & 2.04 (H) & N/A & 2.64 (H) & 2.55 (H) \\
                        & OP & $E_{\mathrm{ads}}$ (kJ/mol) & 22.2 (TT,G) & N/A & 15.9 (TT,E) & 8.4 (TT,E) \\
                        &  & $d_z$ (\AA)                                 & 3.18 & N/A & 3.64 & 4.51 \\
        \hline\hline
     Graphene/ &  OP & $E_{\mathrm{ads}}$ (kJ/mol) & 12.1 (TT) & 11.1 (TT) & 16.9 (TT) & 6.4 (TA) \\
     Benzene                     &  & $d_z$ (\AA)                 & 3.17 & 3.33 & 3.58 & 3.73 \\
     (ring center) & & & & & & \\
        \hline\hline
\end{tabular}
\label{Tab:eads_ch4}
\end{table*}

Despite the quantitative differences, the trends in $E_\mathrm{ads}$ for IP adsorption across the various functional groups are very similar for the three DFT functionals, as well as the MP2 calculations (Fig.~\ref{fig:eads_bymethod}). Interestingly, of the DFT results, the trend for uncorrected DFT-LDA is actually the closest match to the trend for MP2, to within a multiplicative constant (see Fig.~\ref{fig:eads_bymethod}(c)). This finding is consistent with previous results on physisorbed systems, for which DFT-LDA trends compared favorably with competing exchange and correlation functionals \cite{girifalco02, lu09, okamoto01, zhao02} and to Quantum Monte Carlo calculations of weak molecular binding to benzene. \cite{kanai09, kim06} Similarly, DFT-LDA has also been shown to reproduce the correct preferred orientations for methane crystals,\cite{li10} and to provide an accurate description of electron density in weak binding of small molecules to carbon systems.\cite{kanai09, langreth09} The greatest deviations in the computed trends are found in the vdW-DF description of IP binding to the two largest functional groups, COOH and H$_2$PO$_3$ (Fig.~\ref{fig:eads_bymethod}(c)). Binding to these appears relatively weaker than one would expect by comparison with the other DFT methods or with the MP2 calculations. It is unclear if this is an effect of the size of these groups or if it is related to the specific description of the methane binding orientation, which involves simultaneous interaction with an O--H bond and an oxygen lone pair.

For methods other than DFT-LDA, OP adsorption is preferred over IP adsorption. However, the different methods exhibit less variation across the functional groups for OP adsorption than for IP adsorption, making it more difficult to define specific trends. Moreover, a few of the groups did not have stable OP binding sites in certain descriptions, but rather collapsed into the IP configuration or else migrated to a faraway site above the ribbon center. Note that the existence or absence of stable OP binding sites is method dependent; for instance, vdW-DF and MP2 identify a stable OP minimum for OH, whereas DFT-LDA and DFT+LAP do not. This suggests a qualitative dependence of the potential energy surface for OP adsorption on the chosen electronic structure description, which contrasts with what we observe for IP adsorption. An additional feature unique to OP adsorption is the near constancy of the MP2 results across the functional groups. This most likely reflects the differences between small benzene rings and nanoribbons. Nanoribbons can accommodate additional charge reorganization from functionalization through their extended $\pi$ manifold. 

\begin{figure}
\centering
\includegraphics[width=3.0in]{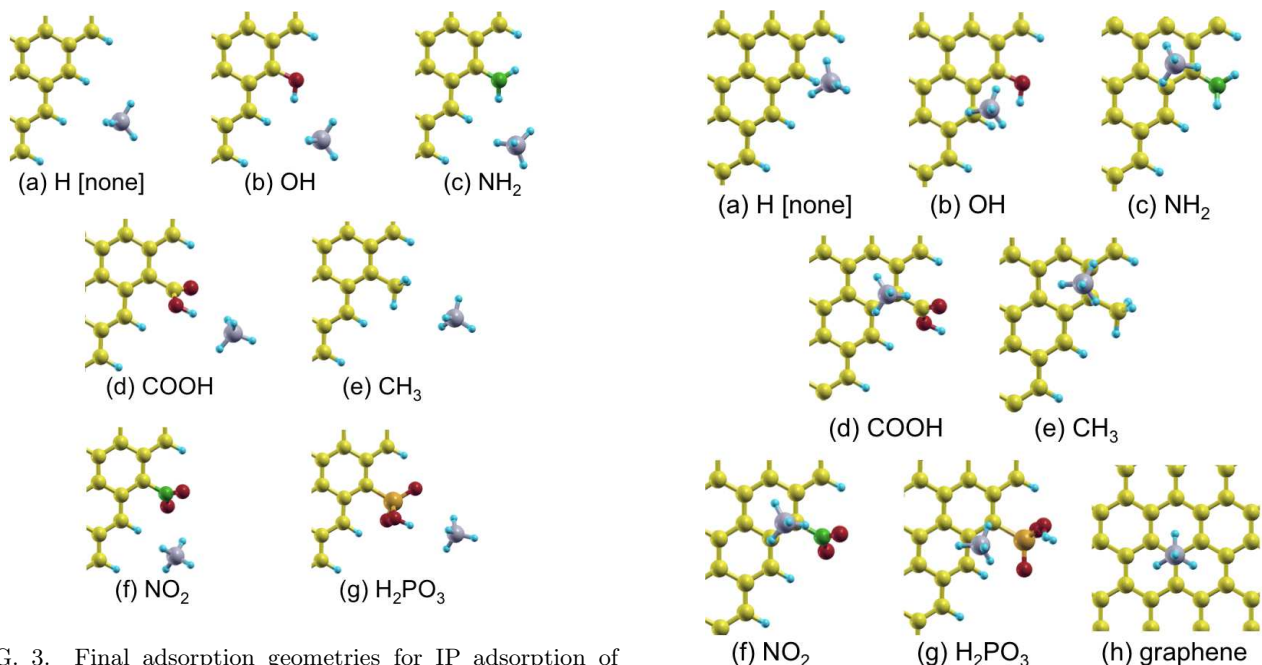}
\caption{Final adsorption geometries for IP adsorption of methane on (a) an unfunctionalized ZGNR; as well as ZGNRs functionalized with (b) OH, (c) NH$_2$, (d) COOH, (e) CH$_3$, (f) NO$_2$, and (g) H$_2$PO$_3$. All shown configurations represent the DFT+LAP result, except the H$_2$PO$_3$-functionalized ZGNR, which represents the vdW-DF result. Additional interactions of the methane with nearby passivation hydrogens are 
evident in (a), (b), (c), and (f). Color scheme: H (cyan), C (yellow), O (red), N (green), P (orange). The methane carbon is shown in gray.
}
\label{fig:ch4_side_geom}
\end{figure}

\begin{figure}
\centering
\includegraphics[width=3.0in]{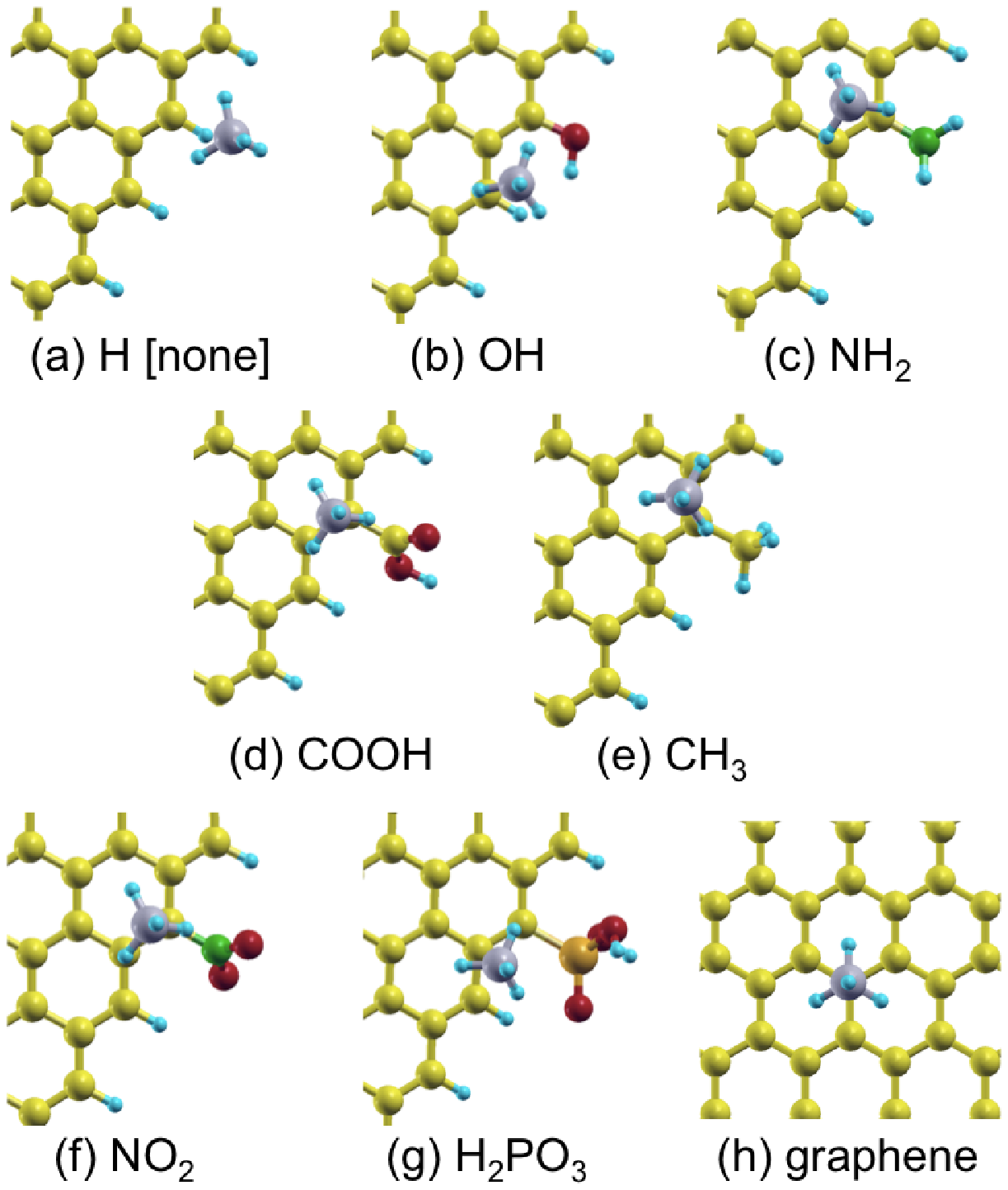}
\caption{Final adsorption geometries for OP adsorption of methane on (a) an unfunctionalized ZGNR; as well as ZGNRs functionalized with (b) OH, (c) NH$_2$, (d) COOH, (e) CH$_3$, (f) NO$_2$, and (g) H$_2$PO$_3$; and (h) on clean graphene. All shown configurations represent the DFT+LAP result, except the H$_2$PO$_3$- and OH-functionalized ZGNRs, which represent the vdW-DF result. Color scheme: H (cyan), C (yellow), O (red), N (green), P (orange). The methane carbon is shown in gray.
}
\label{fig:ch4_top_geom}
\end{figure}

Table~\ref{Tab:eads_ch4} also lists the final orientations of the IP- and OP-adsorbed methane for each of the tested methods and functionalizations. These orientations are depicted in Fig.~\ref{fig:ch4_side_geom} and Fig.~\ref{fig:ch4_top_geom} for DFT+LAP. For both IP and OP adsorption, the orientations are very consistent across the DFT methods, with tripod-toward (TT) by far the most common. The MP2 results are generally compatible with the DFT orientations, although for three of the groups (OH, NO$_2$, and unfunctionalized benzene), the straddle (S) orientation is preferred over the tripod-toward (TT) orientation for IP binding. The availability of nearby carbon rings for direct interaction with methane C--H bonds leads to a TT orientation for methane on graphene, whereas a TA orientation is preferred on benzene.

The specific location of the adsorbed methane for OP binding is also listed in Table~\ref{Tab:eads_ch4} and depicted for DFT+LAP in Fig.~\ref{fig:ch4_top_geom}. The results show that adsorption usually occurs near the zigzag edge carbon to which the functional group is anchored, rather than above the atoms of the functional group itself. The only exceptions to this are the unfunctionalized ZGNR, where DFT-LDA and DFT+LAP show stable binding above the hydrogen edge atoms; and the DFT-LDA result for H$_2$PO$_3$, which shows stable binding approximately above the hydrogen atom in the O--H bond (this is the likely origin of the significant enhancement in $E_\mathrm{ads}$ with respect to the other methods, where a different minimum was found). This suggests that the functional group changes the electronic character of the carbon atom that couples it to the extended $\pi$ system. In addition, the OP geometry features the C--H bond that is farthest from the functional group aligned so as to point toward a ring center (see Fig.~\ref{fig:ch4_top_geom}). This echoes the stable binding configuration for methane on graphene, where each C--H bond orients toward a ring center (Fig.~\ref{fig:ch4_top_geom}(h)).

Comparison of the binding distances for the various methods in Table~\ref{Tab:eads_ch4} reveals that distances for DFT-LDA are generally shortest, followed by DFT+LAP, with vdW-DF the longest. The experimental binding distance of methane to graphite, 3.45\AA, \cite{vidali91} falls between the DFT+LAP (3.33\AA) and vdW-DF (3.58\AA) results for graphene. It is worth noting that based on Table~\ref{Tab:eads_ch4}, binding distances alone cannot be considered a reliable predictor of $E_\mathrm{ads}$.

From the final IP relaxed geometries in Fig.~\ref{fig:ch4_side_geom}, it can be seen that methane aligns primarily along the O--H bond for the OH, COOH, and H$_2$PO$_3$ functionalizations, along the N--H bond for NH$_2$, and along the N--O bond for NO$_2$. The impact of these polar bonds on methane binding geometry suggests significant interaction with the permanent dipole of the functionalized substrate; the additional importance of this quantity in predicting adsorption strength is explored below. In addition, the smaller polar functional groups feature significant secondary interactions with one of the neighboring C--H bonds along the zigzag edge. This is the case for OH, NH$_2$, and NO$_2$. This enables the methane to orient with its tripod toward (TT) the polar bond in the functional group while simultaneously interacting with the side C--H bond in a straddle (S) orientation (these are reversed for NO$_2$, with S and TT interactions for the N--O and C--H bonds, respectively).

The larger COOH and H$_2$PO$_3$ groups feature a doubly bonded oxygen in addition to the hydroxyl group. In these instances, binding is also facilitated by simultaneous interaction of the methane with the O--H bond and the lone-pair site, as seen in Fig.~\ref{fig:ch4_side_geom}(d)\&(f). Examination of the charge density for these groups reveals that binding is accompanied by charge transfer from the lone pairs of the doubly bonded oxygen to the O--H bond of the other oxygen, enhancing the polarity of the O--H bond. The resulting configuration exploits the tetrahedral geometry of the adsorbate to maximize the specific interaction with the functional group.

\section{Carbon dioxide adsorption}

\begin{figure}
\centering
\includegraphics[width=3.0in]{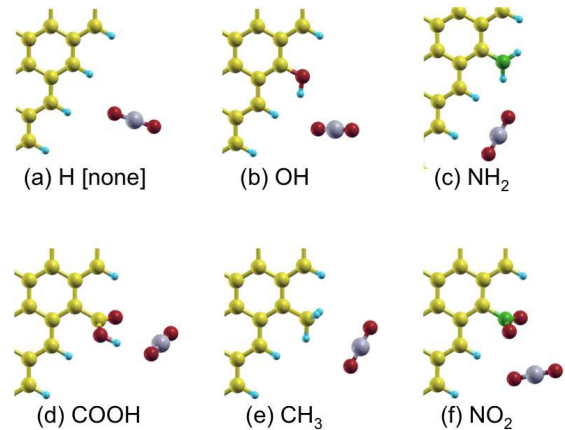}
\caption{Final adsorption geometries for IP adsorption of CO$_2$ on (a) an unfunctionalized ZGNR; as well as ZGNRs functionalized with (b) OH, (c) NH$_2$, (d) COOH, (e) CH$_3$, and (f) NO$_2$. Configurations were obtained using the DFT+LAP method. For the OH- and COOH-functionalized ribbons, the CO$_2$ molecule is rotated about its $C_2$ axis by $\sim 45^\circ$ with respect to the ZGNR plane. Additional interactions of the CO$_2$ molecule with nearby passivation hydrogens are evident in (a), (b), (c), and (f). Color scheme: H (cyan), C (yellow), O (red), N (green). The CO$_2$ carbon is shown in gray.
}
\label{fig:co2_side_geom}
\end{figure}

\begin{figure}
\centering
\includegraphics[width=3.0in]{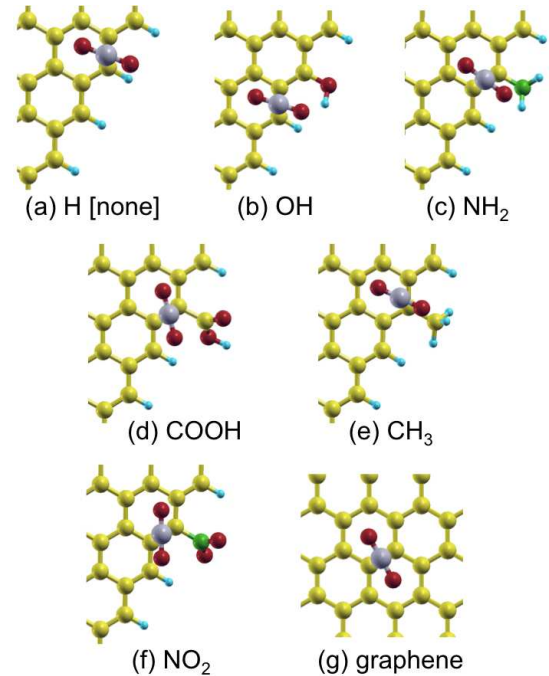}
\caption{Final adsorption geometries for OP adsorption of CO$_2$ on (a) an unfunctionalized ZGNR; as well as ZGNRs functionalized with (b) OH, (c) NH$_2$, (d) COOH, (e) CH$_3$, and (f) NO$_2$; and (g) on clean graphene. Configurations were obtained using the DFT+LAP method. Color scheme: H (cyan), C (yellow), O (red), N (green). The CO$_2$ carbon is shown in gray.
}
\label{fig:co2_top_geom}
\end{figure}

We also calculated binding of CO$_2$ on the functionalized ZGNRs. DFT+LAP was used for these values, since it gave adequate results for methane binding distance, energy, geometry, and trend across functional groups with respect to MP2 and available experimental data. The results for $E_\mathrm{ads}$ and the final relaxed binding geometries are shown in Table~\ref{Tab:eads_co2}, Fig.~\ref{fig:co2_side_geom}, and Fig.~\ref{fig:co2_top_geom}. According to Table~\ref{Tab:eads_co2}, binding of CO$_2$ usually exceeds that of CH$_4$ by about a factor of two, both for the ZGNRs and for clean graphene. This is due primarily to the presence of a quadrupole moment in CO$_2$. The energetic difference between IP and OP adsorption is larger for CO$_2$, with OP adsorption again preferred. Nevertheless, the ordering of $E_\mathrm{ads}$ across the functional groups for both OP and IP adsorption is very similar for both gases. 

A comparison of the adsorption strength of CO$_2$ versus CH$_4$ for a given ZGNR offers insight into the relative selectivity of the substrate (last column of Table~\ref{Tab:eads_co2}). Here, we find that the unfunctionalized (hydrogen-passivated) ZGNR shows an exceptionally high difference between CO$_2$ and CH$_4$ binding strengths: for DFT+LAP, OP (IP) binding of CO$_2$ demonstrates a 257\% (162\%) increase over CH$_4$, compared to the graphene baseline increase of 108\%. This suggests that significantly enhanced selectivity for CO$_2$ over CH$_4$ may be achieved in graphitic systems with high concentrations of hydrogenated exposed edges. This is applicable to natural gas purification, in which efficient, selective CO$_2$ removal from CH$_4$ streams is an important step. Nevertheless, edge concentrations of oxygen- and nitrogen-containing functional groups should be kept sufficiently low, since these groups often decrease selectivity. Moreover, one should keep in mind that thermodynamic competition will favor surface sites over the fully hydrogen-passivated zigzag edge, meaning the realization of any practical benefit would likely require very high edge-carbon concentrations, as well as high gas pressures.

\begin{table}
\centering
\caption{Final relaxed geometries and adsorption energies $E_{\mathrm{ads}}$ of CO$_2$ adsorbed on functionalized and unfunctionalized ZGNRs. The final orientation and position of the CO$_2$ molecule is given alongside the adsorption energy in parentheses (see text for description of acronyms). For IP adsorption, $d_\mathrm{min}$ represents the distance from the CO$_2$ carbon atom to the nearest functional-group atom, whose identity is given in parentheses. For OP adsorption, $d_z$ is the distance of the adsorbate from the ribbon basal plane. For each value of $E_{\mathrm{ads}}$, the last column shows the percent change with respect to the equivalent result for methane adsorption.}
\begin{tabular}
        {ccccc}
        \hline\hline
        Group &  Site & Quantity & DFT+LAP & ~~vs.~CH$_4$ \\
        \hline
         None (H) &  IP & $E_{\mathrm{ads}}$ (kJ/mol) & 11.8 (AT)    & +162\% \\
                        &  & $d_{\mathrm{min}}$ (\AA)           & 3.61 (H)      & \\
                        & OP & $E_{\mathrm{ads}}$ (kJ/mol) & 18.9 (AP,E) & +257\% \\
                        &  & $d_z$ (\AA)                                 & 3.20             & \\
        \hline
         OH          &  IP & $E_{\mathrm{ads}}$ (kJ/mol) & 14.0 (AP)     & +57\% \\
                        &  & $d_{\mathrm{min}}$ (\AA)           & 2.96 (H)       & \\
                        & OP & $E_{\mathrm{ads}}$ (kJ/mol) & 20.8 (AP,E) & N/A \\
                        &  & $d_z$ (\AA)                                 & 3.09             & \\
        \hline
         NH$_2$   &  IP & $E_{\mathrm{ads}}$ (kJ/mol) & 17.8 (AT)    & +82\% \\
                        &  & $d_{\mathrm{min}}$ (\AA)           & 3.34 (H)      & \\
                        & OP & $E_{\mathrm{ads}}$ (kJ/mol) & 25.1 (AP,E) & +96\% \\
                        &  & $d_z$ (\AA)                                 & 3.17            & \\
        \hline
       COOH       &  IP & $E_{\mathrm{ads}}$ (kJ/mol) & 19.9 (AP)    & +101\% \\
                        &  & $d_{\mathrm{min}}$ (\AA)           & 2.78 (H)      & \\
                        & OP & $E_{\mathrm{ads}}$ (kJ/mol) & 21.2 (AP,E) & +86\% \\
                        &  & $d_z$ (\AA)                                 & 3.23            & \\
        \hline
       CH$_3$     &  IP & $E_{\mathrm{ads}}$ (kJ/mol) & 6.0 (AP)      & +114\% \\
                        &  & $d_{\mathrm{min}}$ (\AA)           & 2.98 (H)      & \\
                        & OP & $E_{\mathrm{ads}}$ (kJ/mol) & 22.3 (AP,E) & +125\% \\
                        &  & $d_z$ (\AA)                                 & 3.21            & \\
        \hline
       NO$_2$    &  IP & $E_{\mathrm{ads}}$ (kJ/mol) & 17.0 (AP)     & +87\% \\
                        &  & $d_{\mathrm{min}}$ (\AA)           & 3.03 (O)      & \\
                        & OP & $E_{\mathrm{ads}}$ (kJ/mol) & 20.7 (AP,E) & +82\% \\
                        &  & $d_z$ (\AA)                                 & 3.30             & \\
        \hline\hline
       Graphene &  OP & $E_{\mathrm{ads}}$ (kJ/mol) & 23.1 (AP)   & +108\% \\
                        &  & $d_z$ (\AA)                                  & 3.14           & \\
        \hline\hline
\end{tabular}
\label{Tab:eads_co2}
\end{table}

For IP binding (Fig.~\ref{fig:co2_side_geom}), the specific CO$_2$ orientations vary from group to group. The OH, COOH, CH$_3$, and NO$_2$ ZGNRs favor alignment of the molecular $C_2$ axis with the functional group (AT); the NH$_2$ and unfunctionalized ribbon have the $C_\infty$ axis aligned with the functional group (AP). In addition, most of the groups have the $C_\infty$ axis of the IP CO$_2$ molecule lying parallel to the ZGNR plane. Exceptions are the OH- and COOH-functionalized ZGNRs, for which the $C_\infty$ axis forms a nearly 45$^\circ$ angle with the plane of the ribbon. 

As we saw for the methane case, the ZGNRs functionalized with smaller polar functional groups (OH, NH$_2$, NO$_2$) feature significant secondary adsorbate interactions with neighboring edge hydrogens for IP binding. This is particularly true for NH$_2$, where the AP binding geometry results in the CO$_2$ molecule being centered far along the ribbon axis (Fig.~\ref{fig:co2_side_geom}(c)). Also, the center carbon in the CO$_2$ molecule aligns approximately along the polar bond in the NH$_2$, COOH, and NO$_2$ ZGNRs (Fig.~\ref{fig:co2_side_geom}(c),(d),(f)), again similar to methane. Neighboring edge hydrogens are much more accessible on ZGNRs than on functionalized benzene, due to the latter's ring geometry; accordingly, our geometries for ZGNRs with smaller polar groups differ from those reported in Ref.~\onlinecite{torrisi10} for CO$_2$ on functionalized benzene. 

When looking at OP adsorption on the ZGNRs (Fig.~\ref{fig:co2_top_geom}), the location of the CO$_2$ carbon, as well as its axial orientation, tends to mimic the graphene configuration. On graphene, the CO$_2$ molecule sits parallel to the plane, such that the oxygens lie above the centers of two consecutive rings with the carbon above the center of a C--C bond (Fig.~\ref{fig:co2_top_geom}(g)). Similarly, for OP binding on the functionalized ZGNRs, the CO$_2$ molecule always lies flat in the ZGNR plane, with the CO$_2$ carbon sitting above the ZGNR edges and never above the functional group atoms.  This is even true for the unfunctionalized ZGNR (Fig.~\ref{fig:co2_top_geom}(a)). In addition, the molecule is always centered between two carbons on a zigzag ribbon edge, with one of its oxygens sitting above a ZGNR carbon ring, much like the graphene case. These observations echo our general conclusions for OP methane binding, where the preferred orientation on graphene significantly influenced the final methane configuration. The location of the remaining CO$_2$ oxygen that does not sit above a carbon ring center varies. For NH$_2$, CH$_3$, and the unfunctionalized ZGNR (Fig.~\ref{fig:co2_top_geom}(a),(c),(e)), it shows significant interaction with the functional group (or H, in the case of the unfunctionalized ribbon), resulting in alignment of the $C_\infty$ axis with the primary axis of the group. For other ribbons, direct interaction is generally avoided. This choice of configuration may have some correlation with binding strength, since NH$_2$ and CH$_3$ are also the strongest-binding OP groups (see Table~\ref{Tab:eads_co2}). Note that unlike the methane case, DFT+LAP was able to identify a stable OP binding site for CO$_2$ on the OH-functionalized ZGNR. 

\section{Discussion}

\begin{figure}
\centering
\includegraphics[width=3.3in]{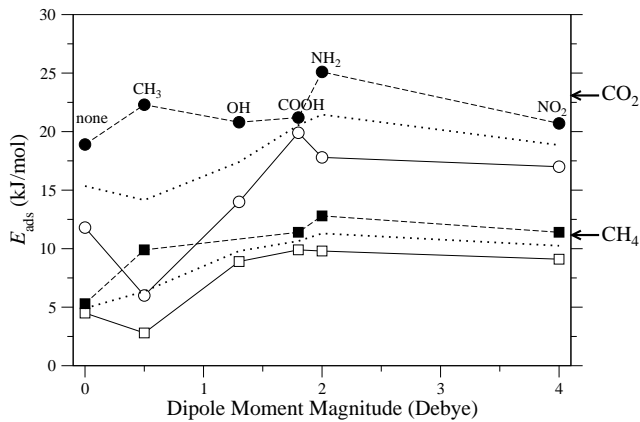}
\caption{Calculated DFT+LAP adsorption energy $E_\mathrm{ads}$ for CH$_4$ (squares) and CO$_2$ (circles) on all ZGNRs as a function of the permanent dipole moment of the correspondingly functionalized benzene substrate. Values for IP (solid line, open symbols) and OP (dashed line, filled symbols) adsorption are shown. The dotted lines track the average of the IP and OP values. Arrows on the right-hand side indicate the calculated values for OP adsorption on a clean graphene sheet.}
\label{eadsvsmu}
\end{figure}

For both IP and OP adsorption, chemical functionalization of the edges can enhance gas binding with respect to the unfunctionalized hydrogen-terminated ZGNR. This can be seen in Tables~\ref{Tab:eads_ch4} and \ref{Tab:eads_co2}, as well as in the summarized results for both gases using DFT+LAP, presented in Fig.~\ref{eadsvsmu}. The calculated degree of enhancement depends on the specific method and functionalization. Nevertheless, increases are universally demonstrated for all functionalizations except IP binding of CH$_3$ when comparing within a given geometry (IP or OP). This suggests that exposed edges in carbon substrates, which are nominally rather inert to gas adsorption when passivated by hydrogen, can become binding sites when properly functionalized. 

For results in Tables~\ref{Tab:eads_ch4} and \ref{Tab:eads_co2} other than uncorrected DFT-LDA, OP binding is clearly preferred over IP binding. For some groups, the differences between the two binding sites are quite large: CH$_3$, for instance, demonstrates relatively strong OP binding in the DFT+LAP and vdW-DF descriptions but is by far the weakest-binding IP group. For others, such as COOH, the binding geometry has only a minor effect on $E_\mathrm{ads}$ in the van der Waals-corrected DFT schemes. However, it is worth emphasizing that even though OP binding seems to be energetically preferred, both geometries are relevant for the understanding and design of nanostructured carbon materials. This is because certain microstructures may preclude one binding geometry or the other. For instance, it is reasonable to assume that graphitic terraces may favor IP binding, so as to permit simultaneous interaction with the terrace-attached functional group as well as the underlying surface layer. Similarly, both IP and OP binding might be exhibited by a single adsorbate at a bilayer step edge, where it could interact simultaneously with the chemical functional group at the edge of one layer via IP binding, and at the edge of the second layer via OP binding. In the case of interactions with functional groups in crystalline nanomaterials such as MOFs or COFs, steric factors could also favor either IP or OP adsorption. \cite{wilmer12}

It should be noted that the observed binding enhancements for the polar functionalizations (especially NH$_2$, NO$_2$, and COOH) over the hydrogen-passivated ZGNR apply only to edge binding. It may be argued that a more apt comparison for application purposes would be to a clean planar graphene surface, against which the associated increases in $E_\mathrm{ads}$ are effectively negligible for results other than uncorrected DFT-LDA (see Fig.~\ref{eadsvsmu}, for example). However, the fact that edge adsorption can be made competitive with surface adsorption also means that edge sites become viable binding sites, thereby enhancing the effective active surface area of the substrate. In nanoporous materials where edges are expected to comprise a significant fraction of available carbon binding sites (e.g., aerogels\cite{kabbour06,biener12}), this should translate to enhanced capacity for gas uptake. Perhaps more significantly, a naive comparison of edge and surface binding belies the importance of the microstructure of the pore walls in nanoporous carbon, where edges and surfaces can be expected to coexist. In these instances, one may assume the effect on the gas binding will be additive, with the functionalized edges providing additional stabilization at the pore wall beyond existent surface-site binding. 

The similarity in the results for CH$_4$ and CO$_2$, despite differences in specific binding orientations and molecular symmetries, points to the importance of the substrate rather than the adsorbate in determining the binding trends. It is reasonable to assume that similar trends could be found for other physisorbed gas species. Note that the agreement between the CH$_4$ and CO$_2$ trends is particularly strong for IP binding, where the direct interaction with the functional group is likely to dominate relative to contributions from the bulk of the ribbon. 

In order to understand the specific origin of the observed trends in $E_\mathrm{ads}$ in the case IP binding, we have carried out a L\"owdin population (LP) analysis on the DFT+LAP results to extract partial charges on the atoms. Changes in these quantities can be tracked to estimate the charge redistribution induced by adsorption, both for the functional group and for the adsorbate. We point out that charges derived from an LP analysis are meaningful only in a relative sense; we use them as an approximate indicator of correlation between induced charges and observed adsorption trends. The LP charges induced upon IP adsorption of CH$_4$ on the functionalized ZGNRs (OH, NH$_2$, COOH, CH$_3$, and NO$_2$) are listed in Tables~\ref{Func_Lowdin} and \ref{Methane_Lowdin} for the functional group and the methane molecule, respectively. As expected for weak-binding systems, the magnitudes of the induced charges are small. The LP analysis shows that the charge distribution of the methane distorts slightly upon adsorption, resulting in induced methane multipole moments that interact electrostatically with the functionalized ZGNR (see Table~\ref{Methane_Lowdin}). The magnitudes of the charges show that methane charge asymmetry is significantly larger for COOH, OH, NH$_2$, and NO$_2$ than for CH$_3$, as expected based on the values of $E_\mathrm{ads}$ in Table~\ref{Tab:eads_ch4}. Also, in each of the stronger-binding cases, one of the methane C--H bonds is significantly more polarized than the others. Close examination reveals that for the OH, NH$_2$, and COOH groups, this always involves the hydrogen atom that is farthest from the functional group, whereas for the strongly electron-withdrawing NO$_2$ group, it involves the hydrogen that is closest to the group. The induced LP charges on the functional group atoms, shown in Table~\ref{Func_Lowdin}, show similar results, with larger changes occurring for COOH, NH$_2$, and OH than for the weak-binding CH$_3$. The calculations in Table~\ref{Func_Lowdin} also demonstrate that the largest changes are found for atoms within the polar O--H and N--H bonds of the COOH, OH, and NH$_2$ groups, further confirming that these polar bonds are primary contributors to methane binding. However, the NO$_2$ group is a notable exception to these trends, exhibiting very weak induced LP charges on the functional group despite a rather large $E_\mathrm{ads}$ and significant induced methane polarization. This could simply be an artifact of the LP partitioning scheme---for instance, no LP change would be observed if the charge redistribution remains largely localized to the oxygen lone pairs. For this reason, it may be desirable to use a partitioning method that has a rigorous connection to the induced polarization, such as a methodology based on maximally localized Wannier functions.\cite{marzari97}

\begin{table}
\centering
\caption{Induced charges (units of $10^{-3}e$) on the functional-group atoms after IP methane adsorption on a ZGNR, based on LP analysis. Results are obtained using the DFT+LAP method. (O1,H1) refers to pairs of atoms in O--H bonds.}
\begin{tabular}{cccccc}
\hline
      &   OH  &  NH$_2$  &  COOH  &  CH$_3$  &  NO$_2$ \\
\hline
  C           &         &          & -3.0    & +2.7 &       \\     
  O1         & +4.5 &          & +10.8 &         & -1.3 \\    
  O2 (=O) &         &          & -3.5   &         & -1.0 \\    
  N           &         & -4.5   &           &        & -4.5 \\    
  H1         & -8.9 & -8.4    & -8.1   & -3.3 &         \\   
  H2         &        &  -13.9 &           & -2.8 &        \\   
  H3         &        &           &           & -2.8 &        \\   
 \hline
\end{tabular}
\label{Func_Lowdin}
\end{table}

\begin{table}
\centering
\caption{Induced charges (units of $10^{-3}e$) on atoms within methane after IP adsorption on a functionalized ZGNR, based on LP analysis. Results are obtained using the DFT+LAP method.}
\begin{tabular}{cccccc}
\hline
   & OH & NH$_2$ & COOH & CH$_3$ & NO$_2$ \\
\hline
  C    & -9.9    & -7.5    & -18.1  & +2.0 & -1.9 \\
  H1  & +13.4 & +10.4 & +12.6 & +1.7 & +3.0 \\
  H2  & +3.7   & -3.1    & +7.7   & -3.6 & +1.4 \\
  H3  & -6.8    & -3.5    & -4.8    & -4.2 & -7.1 \\
  H4  & -6.3    & -5.6    & -5.0    & -4.5 & -14.4 \\
\hline
\end{tabular}
\label{Methane_Lowdin}
\end{table}

Although the LP analysis permits estimation of methane dipole moments, it is more difficult to extract a measure of the polarization induced on the extended ZGNRs upon adsorption. This is because of technical issues that arise when computing dipole moments of extended structures, given that electronic polarization along the repeating directions is ill defined within periodic boundary conditions.\cite{martin74} Alternatively, if we can assume that the dominant interactions between the methane and the substrate are local, we can use functionalized benzene molecules to estimate the dipole moments induced on the functionalized ZGNRs. Since the molecular systems are isolated, there are no complications in computing dipole moments. In Fig.~\ref{eadsvsmu}, we plot the DFT+LAP $E_\mathrm{ads}$ for IP and OP gas binding on functionalized ZGNRs as a function of the magnitude of the permanent dipole moment of the corresponding benzene systems in the absence of an adsorbate, as calculated from DFT+LAP benzene models. Note that since an unfunctionalized benzene molecule possesses no dipole moment, any exhibited polarization arises solely from the presence of the functional groups. Although OP and IP binding show somewhat different behavior, the average of $E_\mathrm{ads}$ for the two geometries correlates with the dipole moment up to $\sim$2~Debye. $E_\mathrm{ads}$ saturates beyond this point, with the highly polar NO$_2$ group showing no additional enhancement. Note that there is no obvious dependence of $E_\mathrm{ads}$ on whether the attached group is electron withdrawing or electron donating in nature. In other words, the physical amount of charge present in the $\pi$ system is not as important for determining adsorption strength as is its arrangement, manifest in the induced polarization.

\section{Conclusions}

In summary, we have shown that the binding of methane to graphitic carbon edges can be significantly modified by the presence of chemical functional groups. The degree of modification is found to be a function not only of the chemical group but also of the gas adsorption geometry. Polar groups, including COOH, NH$_2$, NO$_2$, and H$_2$PO$_3$ are found to be promising candidates for enhancing CO$_2$ and CH$_4$ capacity by activating exposed edges to introduce additional binding sites. This has important implications for improving uptake in carbon structures that are known to have high quantities of edge carbons, including nanoporous activated complexes, carbon aerogels, and crystalline covalent framework materials.

In contrast, by far the highest selectivity for CO$_2$ over CH$_4$ was found for the unfunctionalized, hydrogen-passivated zigzag edge. In this particular case, the CO$_2$ binding preference was significantly higher than for bulk graphene. This implies that higher edge concentrations could be beneficial for natural gas purification using nanoporous carbon, provided the substrates contain relatively low concentrations of other functional groups. Nevertheless, since the hydrogen-passivated zigzag edge was also among the weakest adsorbers, combining high selectivity with high uptake for maximum separation effectiveness remains a design challenge.

The observed trends in binding energies with functional group identity are largely preserved across the DFT-LDA, DFT+LAP, and vdW-DF formulations of the DFT exchange-correlation functional if one considers only in-plane or only out-of-plane binding sites. Nevertheless, comparisons between in-plane and out-of-plane binding energies are not qualitatively consistent across the methods, suggesting that one should use caution in directly comparing results based on very different binding geometries. 

We expect that our work will stimulate experimental efforts to confirm our results, by processing activated carbons to favor the presence of these chemical functional groups and looking for a concomitant change in methane uptake. This should constitute a realistic step towards tailoring activated carbons for adsorptive on-board vehicular natural gas storage. Furthermore, our quantitative comparison of the impact of functionalization on weak gas interactions has relevance for devising chemical modification strategies for carbon-based gas sensing, as well as selective materials for gas separation. 

\section{Acknowledgments}

Computations were performed using the facilities of the Centre for Computational Materials Science, Jawaharlal Nehru Centre for Advanced Scientific Research, Jakkur, India. Funding from Bharat Petroleum Corporation Limited (BPCL), India, is gratefully acknowledged. BW acknowledges funding from U.S. National Science Foundation Grant 701180. A portion of this work was performed under the auspices of the U.S. Department of Energy by Lawrence Livermore National Laboratory under Contract DE-AC52-07NA27344. We also acknowledge useful discussions with G.~Vasudev, N.~V.~Choudary and B.~Newalkar of BPCL, India. We thank Y.~Kanai of the University of North Carolina, Chapel Hill for providing the DFT+LAP pseudopotentials.

\end{document}